**Title:** Development of an updated, comprehensive food composition database for Australian-grown horticultural commodities

**Authors:** Eleanor Dunlop [1,2†], Judy Cunningham [1†], Paul Adorno [3], Shari Fatupaito [4], Stuart K Johnson [5], Lucinda J Black [1,2]

**Author affiliations:**

[1] Curtin School of Population Health, Curtin University, Kent Street, Bentley WA 6102, Australia. eleanor.dunlop@curtin.edu.au; judyc121@gmail.com; lucinda.black@curtin.edu.au

[2] Institute for Physical Activity and Nutrition (IPAN), School of Exercise and Nutrition Sciences, Deakin University. e.dunlop@deakin.edu.au; lucinda.black@deakin.edu.au

[3] National Measurement Institute, 1/153 Bertie Street, Port Melbourne, VIC 3207, Australia. Paul.Adorno@measurement.gov.au

[4] Food Standards Australia New Zealand, 15 Lancaster Place, Majura Park, ACT 2609. shari.fatupaito@foodstandards.gov.au

[5] Ingredients by Design, Lesmurdie, WA 6076. stuart@ingredientsbydesign.com.au

[†] Equal contribution

***Corresponding author:** Lucinda J Black, Institute for Physical Activity and Nutrition (IPAN), School of Exercise and Nutrition Sciences, Deakin University, 221 Burwood Highway, Burwood, VIC 3125, Australia. lucinda.black@deakin.edu.au. Tel.: +61 3 924 45491.

**Financial support:** This work was supported by Hort Innovation (ST19036). Role of the Funder: Authors worked with Hort Innovation to develop the scope for this project and to identify commodities of priority for inclusion. Authors wrote the sampling and analysis plans, which were approved by Hort Innovation. Hort Innovation had no role in the conduct of the study, preparation and interpretation of the study's findings or writing of this article.

**Running title:** Updated food composition data for Australian horticultural commodities

**Abstract**

Australian agriculture supplies many horticultural commodities to domestic and international markets; however, food composition data for many commodities are outdated or unavailable. We produced an up-to-date, nationally representative dataset of up to 148 nutrients and related components in 92 Australian-grown fruit (fresh n=39, dried n=6), vegetables (n=43) and nuts (n=4) by replacing outdated data (pre-2000), confirming concentrations of important nutrients and retaining relevant existing data. Primary samples (*n* = 902) were purchased during peak growing season in Sydney, Melbourne and Perth between June 2021 and May 2022. While new data reflect current growing practices, varieties, climate and analytical methods, few notable differences were found between old and new data where methods of analysis are comparable. The new data will be incorporated into the Australian Food Composition Database, allowing free online access to stakeholders. The approach used could serve as a model for cost-effective updates of national food composition databases worldwide.

**Keywords:** Australia; food composition; fruit; horticulture; nutrient, nuts; vegetable

## 1. Introduction

The Australian horticulture sector is a multi-billion dollar industry that supplies domestic and international markets with a wide range of produce (Hort Innovation, 2023). Plant foods, even in their natural, unaltered state, offer the attraction of being nutrient dense. This attribute should be highlighted to increase their marketability and consumption, leading to economic benefits to food producers and manufacturers, along with health benefits to consumers. However, up-to-date and accurate food composition data are needed to identify key nutrients and enable marketing and promotion of the commodities that contain them.

The majority of data reported in the Australian Food Composition Database (AFCD) for many fresh fruits and vegetables (e.g., potatoes, silverbeet, cherries, papaya, lemon, cos lettuce, peas, plums) were generated over 30 years ago (Food Standards Australia New Zealand [FSANZ], 2019d). For example, much AFCD data for potatoes date back to the 1980s (FSANZ, 2019d), and the potato varieties analysed do not represent all varieties that are now commonly available to consumers. The AFCD aims to include a minimum dataset of 60 nutrient components per food. While the majority of data in the AFCD is the result of direct analysis of composite samples of the identified foods, some data are imputed from similar foods or are borrowed from food composition tables in other countries.

Nutrient content can vary by factors that may change over time, such as geographical or climatic conditions, production and storage methods and variety (Greenfield & Southgate, 2003). Additionally, analytical methods have developed and improved over time, allowing for more accurate and precise determination of nutrient profiles across a wider range of nutrients. Data in the AFCD for a limited number of widely consumed plant foods have been updated recently as part of small analytical programs (FSANZ, 2019a, 2019b, 2019c; 2019d);

however, many data remain outdated, missing, imputed or borrowed. This is not surprising given the cost of generating new, representative food composition data for the wide range of nutrient components included in food composition tables (typically exceeding USD$3000 per food, including vitamins, minerals, fatty acids and proximate components, excluding sampling cost).

Hence, this project generated new food composition data to replace outdated data, and to fill data gaps for Australian-grown horticultural commodities, in a cost-effective manner. This dataset was to be of suitable quality and scope to be incorporated into the AFCD and to support industry marketing and educational materials. In this study, we aimed to assess the magnitude of any differences between existing and new data, as a guide to decision-making on the use of older data into the future.

## 2. Methods

### *2.1 Identifying outdated and missing data*

The total project budget, while considerable, did not allow complete re-analysis of all Australian-grown commodities available on the domestic market. Therefore, an approach to selection of commodities and components was developed as follows. Horticultural commodities were initially selected for inclusion in the project through consultation with Hort Innovation, a not-for-profit horticultural research and development corporation that is owned by Australian levy-paying horticultural growers. Initial review of the AFCD data identified almost 200 data lines for raw fruits, vegetables, nuts and dried fruits, some of which represent commodities imported into Australia, not relevant to the funding source and/or of limited consumption in Australia. At this stage, several “new” fruits and vegetables grown in Australia were identified that had not previously been analysed or included in the AFCD, including new colour variants of existing commodities. From this review, and in consultation with Food Standards Australia New Zealand (FSANZ), the managers of the AFCD, a total of 92 foods (39 fresh fruits, 6 dried fruits, 43 fresh vegetables and 4 nuts) were selected for analysis.

For the selected foods, gaps in nutrient data and outdated nutrient data were identified from AFCD supporting materials (FSANZ, 2019d). Outdated values were considered to be those:

- that were generated before the year 2000, regardless of the analytical method used
- that were generated after the year 2000 but the methods of analysis have been modified and greatly improved in accuracy and precision (e.g., folates)
- where the commodity itself has undergone major varietal change, for example widespread availability of differently coloured types.

Missing components were those that:

- were not included at the time the current data were generated (e.g., folates, iodine) but may be present in quantifiable amounts
- were values borrowed from another country’s food composition tables
- were values imputed from a related Australian food.

Foods or specific components were also included, in some cases, if there was industry interest in making marketing statements based on content of these components, such as lutein, to provide certainty as to the basis of the claim. Some components were omitted from analysis where analysis cost was high and the food was not expected to contain nutritionally relevant levels in relation to the Australia New Zealand Nutrient Reference Values ((National Health and Medical Research Council, 2020) (e.g., vitamin D, for which data were published recently for relevant foods in Australia (Dunlop et al., 2021)). For this purpose, nutritional relevance was considered to be more than 10% of the Recommended Dietary Intake or Adequate Intake for adults per 100 g edible portion.

*2.2 Sample purchasing*

A sampling schedule was developed across 12 sampling rounds that occured monthly from June 2021 to May 2022. Commodities were assigned to a sampling round according to their growing season with the aim of measuring nutrient content at the peak of the growing season. Australian-grown samples were purchased by trained National Measurement Institute (NMI) staff according to detailed guidelines. Each sample was labelled with a unique NMI sample ID, the commodity type and the state of origin. The store name and location and date of purchase were recorded. Commodities were purchased across three cities (Sydney, Melbourne and Perth) that represent the east and west coasts of the continent and are where approximately half of the Australian population resides. Samples were couriered to NMI's food laboratory in Port Melbourne and remained chilled from point of purchase until preparation.

For each commodity, the aim was to collect 10 primary samples, with a minimum of 7 samples. Where this was not achieved in the intended sampling phase, additional primary samples (dried apricot, Brussels sprouts, white cabbage and white onions) were collected at the end of the project and analysed separately, with the exception of persimmon where none were available at that time. Three commodities (mandarin, orange sweet potato and passionfruit) were re-purchased at the end of the project to confirm specific results that had been queried during a review of the data.

*2.3 Sample preparation*

Primary samples were photographed upon arrival at the laboratory in order to capture the NMI sample ID, colour, ripeness, condition, variety and all aspects of product packaging and produce stickers, where applicable. Samples were prepared (e.g., peeled, trimmed, de-seeded) as they would usually be consumed.

For 37 commodities likely to contain substantial levels of vitamin C, an aliquot of each primary sample was taken for immediate individual analysis of vitamin C in order to minimise preparation time and destruction of the nutrient. For all other nutrients, an homogenised composite sample was created for each commodity using equal aliquots of the primary samples. The majority of samples were analysed raw. Chestnuts were analysed only in the cooked form (they are not consumed in the raw form) and eggplant and mushroom were analysed in both cooked and raw forms to confirm selected nutrient changes on cooking. Chestnuts were weighed and baked (18 minutes at 200°C) whole with shell on, and were turned halfway through cooking. Mushrooms were weighed and pan fried without added oil or fat until they turned a light brown colour and appeared well-cooked. Eggplant were weighed and baked (30 minutes at 180°C) until the flesh and skin appeared soft/collapsed and well-cooked. Included commodities are detailed in Supplementary Table 1.

*2.4 Quality control*

Analytical quality control was monitored at multiple times during the 11 month study. Monitoring included:

- determination of relative percent differences (RPD) between replicates. RPDs of 5-15% were regarded as acceptable where results were greater than 10 times the Limit of Reporting (LOR);
- comparison against standard reference materials (NIST1546a meat homogenate; FAPAS 25172 Corn/Maize snack; FAPAS 21113 Breakfast cereal) where such materials were available for the nutrients being analysed. In-house reference materials (cocoa powder, apricot jam) were also used where relevant. All results were within acceptable ranges;
- Spike recoveries. Recoveries of 90-100% were considered acceptable and achieved for all analytes tested, other than tryptophan (minimum 87% recovery) and lutein (115% in one sample).

*2.5 Analysis*

Analyses were conducted at NMI, except: dietary fibre and starch (Australian Export Grains Innovation Centre [AEGIC], North Ryde), vitamin B5 (Bureau Veritas AssureQuality [BVAQ], Melbourne and NMI), vitamin B7 (BVAQ, Melbourne) and folate (PathWest, Perth) (Supplementary Table 2). All laboratories involved in the analyses were accredited (ISO17025) by the National Association of Testing Authorities for analysis of the respective nutrients, with the exception of vitamin B5 measured at NMI. In January 2022, technical issues with VitaFast test kits forced discontinuation of analysis of vitamin B5 at BVAQ; therefore, NMI conducted analysis of vitamin B5 from February 2022 using a quadrupole time-of-flight and quattro micro tandem mass spectrometry method based on earlier methods (AOAC International, 2012; Mittermayr, Kalman, Trisconi, & Heudi, 2004; Pakin, Bergaentzlé, Hubscher, Aoudé-Werner, & Hasselmann, 2004). Samples analysed in January 2022 at BVAQ were retested using NMI's vitamin B5 method to confirm that similar results were returned by the two methods. As the reanalysis confirmed the initial results, the mean value of both analyses was reported.

*2.6 Data handling and quality checks*

All data were reviewed by two team members (JC and ED) to identify factors such as completeness, correct units and unexpected results. Three commodities with unexpected results (folates in mandarins, carotenes in passionfruit and tocopherols in orange sweet potato) were subject to reanalysis and the mean value reported unless an issue was identified in the original data.

*2.7 Statistical analysis*

As the majority of data points were derived from composite samples, limited statistical analysis was undertaken. For commodities with individual ascorbic acid values, standard deviation of the sample was determined using the STDDEV.S function of Microsoft Excel™. For the purposes of generating mean values, any new values determined as being below the LOR were assigned a half LOR value, unless all included values were <LOR.

*2.8 Development of the comprehensive dataset*

As components were primarily included in this study where they were missing or outdated in the AFCD, it was necessary to combine existing AFCD data with our new data to develop the full comprehensive dataset. Where appropriate, some values were imputed from related commodities analysed in this study (e.g., the protein and fat content of red cabbage was imputed from the analysed values for white cabbage). Existing and imputed data were incorporated after an adjustment to express values on the dry matter basis found in this study.

$$Dry\ matter\ adjusted\ value = \frac{Existing\ AFCD\ value}{(100 - Existing\ moisture\ content)} * (100 - New\ moisture\ content)$$

Values for energy, total sugars, niacin derived from tryptophan, niacin equivalents, vitamin A equivalents, and total saturated-, monounsaturated- and polyunsaturated fatty acids were then calculated (Food and Agriculture Organization of the United Nations [FAO] & International Network of Food Data Systems [INFOODS], 2012). From this, nutrient values for selected cooked vegetables were calculated using established recipe calculation techniques (FAO, 2021). The comprehensive dataset has been incorporated into FSANZ’ data holdings for publication within future releases of the AFCD.

## 3. Results and discussion

### *3.1 Analytical and comprehensive datasets*

A total of 3744 component values were generated across these 92 foods from 902 primary samples. The majority were vitamins and related compounds, and minerals.

The comprehensive dataset contained nearly 9000 values, including over 3700 data points generated by direct analysis, as well as recipe-calculated values for 56 foods covering boiled, baked and/or fried (without fat or oil) vegetables that are usually consumed in the cooked form.

### *3.2 Change in values over time*

Analysed values for key nutrients were compared to existing values in AFCD in Table 1, for those commodities with existing AFCD data.

Across most commodities and components, there was little substantial change in nutrient values over time, with the majority of new values falling within 20% of the currently published values. Moisture and sugars were more consistent across time than the remaining components. This was also the case for the subset of commodities analysed at two different time points at least 6 months apart under the same general sampling considerations. Due to the consistency in moisture content, in particular, values for remaining nutrients were not calculated and expressed on a dry matter basis.

Across the four commodity groups, total folates showed consistent increases in values, from a 19% increase in fresh fruits to a >100% increase in dried fruits and nuts. Of the six commodities with starch contents above 1 g/100 g, and where existing data was generated in Australia (i.e., not borrowed or imputed from another source), five showed substantial increases in folates, for example from 57 to 104 µg/100 g in green peas. Levels were largely constant in Cavendish bananas, but the existing values for those were determined after triple-enzyme extraction was introduced.

For fresh fruits, zinc was lower in new data for fresh fruits (28% lower, 0.05 mg/100 g; less than 1% of the recommended daily intake (RDI) for adult females) and copper was lower in dried fruits (27% lower, 0.12 mg/100 g; 10% RDI), overall. For nuts, alpha tocopherol was lower (28%, 2.5 mg/100 g; 35% RDI), as was beta carotene (2.5 µg/100 g; <1% RDI) compared to existing data. Although ascorbic acid was higher in new data for nuts, this was driven by one sample (chestnuts).

For fresh vegetables, in addition to overall increases in folates, alpha tocopherol increased by 28% on average (0.2 mg/100 g), representing only a 3% increase compared to the adult female RDI (7 mg/day). In contrast, decreases of more than 20% from existing values were noted for beta carotene (30%, 354 µg/100 g; approximately 8% of the vitamin A RDI), ascorbic acid (22%, 11 mg/100 g; approximately 25% of the RDI) and iron (26%, 0.2 mg/100 g; less than 1% of the adult male RDI).

Data for minerals included in this study were determined by atomic absorption spectroscopy (AAS) until the mid-1980s, as used in the work reported by Lim (1985). After this time they were generated using inductively coupled plasma mass spectroscopy (ICP-MS). Mean iron and zinc contents of 24 fruits and vegetables generated using AAS were compared with the data from our study for the same commodities. On average, iron was 33% below the older value and zinc was 28% below the older value. In contrast, when the data from our study was compared with 28 commodities analysed using ICP-MS for both old and new data, iron was only 11% lower and zinc 6% lower than in the older data, suggesting the possibility of a methodological effect on reported concentrations of iron and zinc.

Ascorbic acid levels were measured in individual samples for 37 of the commodities in our study. Table 2 summarises the findings and shows high variability within samples of the same commodity purchased at the same time point. Relative standard deviation ranged from 23 to 66% of the mean value.

Nutritionally relevant changes in mean values for carotenes and ascorbic acid were noted in some commodities. Vitamin C was lower overall than in older data, which may be a result of a move away from microfluorometry to automated liquid chromatographic methods (Wills, Wimilasiri, & Greenfield, 1983). Older carotenes data were typically generated using manual chromatographic techniques and the lower values in our study may reflect a move to automated chromatographic systems and improved separation and quantitation of different types of carotenes. Nevertheless, these differences could also reflect changes to market practices such as harvesting at an earlier ripeness with less developed colour, as well as changes in varieties grown and in storage conditions at retail level (Lee & Kader, 2000).

There have been reports of apparent reductions in minerals such as iron, zinc and copper in fresh produce in recent years (Colino, 2022; Eberl, Li, Zheng, Cunningham, & Rangan, 2022) Colino, 2022). This study showed decreases of more than 20% compared to existing AFCD values in iron (vegetables) and zinc (fruits), but not in copper. The magnitude of reductions were small (around 2%) in relation to Nutrient Reference Values. It is not possible to conclude whether or not the decreases represent true reductions or are a result of different analysis methods (Jajda, Patel, Patel, Solanki, Patel, & Singh, 2015) and varieties (Marles, 2017) or simply an artefact of inadequate sample size for components that are highly variable. The finding of greater apparent reductions in iron and zinc when older data was generated using AAS, compared to the current ICP-MS technique, suggests that it would be wise to update such data in food composition tables in future programs.

The question of whether or not nutrient levels have changed over time is challenging to determine in the absence of sufficient information on variability in concentrations of individual nutrients, determined using comparable methods of analysis, at different time points. An examination of the variability in ascorbic acid levels in our study, and a comparison of the same commodities purchased at different times during our study, reinforces the findings of other studies (Arivalagan et al., 2012; Beltramo, Bast, Dilien, & de Boer, 2023; Lim, 1985; Strålsjo, Cornelia, Witthof, Sjoholm, & Jagerstad, 2003) that show large variation in micronutrient levels in retail samples of fruits and vegetables, within and between countries. For some vitamins and minerals, where apparently large changes in concentration were seen, it is not possible from our study to determine the cause of the change or, indeed, to be confident that true changes are being observed that are independent of wide variation in nutrient concentrations within the same commodity.

*3.3 New foods included*

Thirteen foods not previously included in the AFCD were analysed in this study, the majority representing colour variants of existing plants: pink grapefruit, white-fleshed nectarines, peaches and sweet potatoes, red-fleshed papaya, yellow-fleshed capsicum, and purple-fleshed sweet potato, as well as the Kent pumpkin and Shepard avocado varieties. Peeled versions of red-skinned apples and unpeeled versions of orange sweet potato and green-skinned cucumber were included. The majority of differences between new and existing varieties were in components relating to colour, most notably in beta carotene, which was 4.5 times higher in red compared to yellow papaya and lutein, which was 51 times higher in Kent pumpkins compared to Butternut pumpkins. In avocadoes, there was a large difference in folates (1.7 times higher in Shepard compared to Hass), levels of which may be associated with green colours. The only important differences in proximate components were for sweet potatoes, where higher moisture levels in orange sweet potato were accompanied by lower starch content compared to purple and white sweet potato. For those commodities where both a peeled and unpeeled sample was analysed, the main difference was in copper for apples, suggesting a concentration of colour compounds in the peel. The effect of peeling on dietary fibre was small.

*3.4 Considerations when developing and updating food composition databases*

The relative stability of many nutrient values in the included fresh commodities, compared to data generated up to more than 30 years ago, suggests that widescale repeat analyses of nutrients in fresh fruits and vegetables may not be necessary for all data that is to be used in food composition tables and for related activities. It also suggests that discarding old data may not be warranted unless specific conditions apply, such as changes in varieties available and major changes in analytical methodology. In particular, the lack of variation of proximate components such as moisture and sugars suggests that these may not need to be repeated at all unless, for example, a new variety is specifically bred to be sweeter or, as for sweet potatoes, to be less starchy, and where moisture content is needed to determine concentrations on a dry matter basis. Initial consultation with producers will help to identify where such changes are likely to have occurred.

As shown in this study, with several new colour variants analysed for the first time, if funds are limited, analysis should focus on components reflecting colour, such as carotenes and folates, which were found to be the the main differences between varieties of the same plant food. In addition, in future updates to data for food composition tables, it would be worthwhile to include new components not already included in tables. For example in this study, components such as lutein and lycopene were found in high levels in some commodities.

Where major changes in analytical methodology are known to have occurred, such as the move to triple enzyme (amylase, lipase and protease) sample treatment to release bound folates, rather than measurement of undecongugated folates (Iwatani, Arcot, & Shrestha, 2003), this study has shown the importance of considering whether such changes are likely to impact the foods being investigated. For example, we found total folates were higher in starchy fruits, vegetables and nuts, than in older data. In such cases, re-analysis and replacement of old data with new data may be warranted.

The sample size used in this study, and in other Australian food composition studies, is likely to be insufficient to generate a robust mean value. In this study, the relative standard deviation overall within commodities was around 35% for ascorbic acid. For a 90% accuracy of the mean value, likely sample size for a robust estimate of ascorbic acid concentration in citrus fruits is at least 30 samples (Greenfield et al., 2003). For highly variable components, such as vitamins, the best approach for food composition tables and use of the data for marketing and education may be to combine new and old datasets to achieve a better estimate of mean concentrations by increasing the number of data points on which they are based, assuming comparable methods of analysis and no major changes in the commodity itself.

We generated an extensive dataset that can be used to update and extend existing nutrient composition holdings for Australian horticultural products. It has shown that age of data alone is not the only factor in deciding whether or not data should be replaced. For future studies in this group of commodities, if funds are limited, research should focus on vitamin-like components that reflect colour and ripeness, as these were the areas where the greatest differences from existing data were found. Nutrients where there has been a potentially major changes in analytical methodology, such as for folates and minerals, should also be a high priority for updates. In addition, for highly variable nutrients that are present in nutritionally relevant amounts, such as vitamin C, it would be worthwhile to aim for a larger sample size in order to generate a more robust estimate of the mean value, although achieving this within budget constraints will inevitably reduce resources that can be applied elsewhere. Given this, researchers and managers of food composition tables should attempt to educate data users about the uncertainty in food composition datasets and to provide estimates of standard deviation, or similar.

**Authorship: Eleanor Dunlop:** Conceptualization, Funding acquisition, Methodology, Formal analysis, Data curation, Writing – Original Draft, Project administration **Judy Cunningham:** Conceptualization, Funding acquisition, Methodology, Formal analysis, Data curation, Writing – Original Draft **Paul Adorno:** Funding acquisition, Methodology, Resources, Investigation, Writing – Review and Editing **Shari Fatupaito:** Methodology, Writing – Review & Editing **Stuart Johnson:** Methodology, Funding acquisition, Data curation, Project administration, Writing – Review & Editing **Lucinda Black:** Conceptualization, Methodology, Funding acquisition, Writing – Review & Editing, Supervision.

**Acknowledgments:** We extend our heartfelt thanks to Ms Anna Vincent for her contribution to gap analysis of data in the AFCD and responses to queries relating to the AFCD.

**Conflict of interest:** Hort Innovation was involved in study design (selection of commodities for analysis and approval of the sampling plan and nutrient by commodity matrix). Hort Innovation had no role in interpretation or reporting of data, or in writing of this manuscript of the decision to submit for publication.

**Table 1.** Comparison of existing nutrient values in the Australian Food Composition Database (AFCD) for fresh and dried fruits, vegetables and nuts (a) to those determined in this study (b). All values are per 100 g edible portion, fresh weight. [1]

| **Commodity name** | **Moisture (g)** | | **Sugars, total (g)** | | **Dietary fibre (g)** | | **B-carotene (μg)** | | **Ascorbic acid (mg)** | | **Folates, total (μg)** | | **Iron (mg)** | | **Zinc (mg)** | | **Copper (mg)** | |
|---|---|---|---|---|---|---|---|---|---|---|---|---|---|---|---|---|---|---|
| | a | b | a | b | a | b | a | b | a | b | a | b | a | b | a | b | a | b |
| ***Fruits*** [2] | | | | | | | | | | | | | | | | | | |
| Average fresh fruit | 84.5 | 84.7 | 9.5 | 9.5 | 2.8 | 2.3 | 200 | 200 | 22 | 20 | 29 | 34 | 0.27 | 0.23 | 0.18 | 0.13 | 0.068 | *0.065* |
| Percentage difference from existing value, average | 0.2 | | 3.9 | | -17.3 | | 0.1 | | -9.1 | | 18.8 | | -14.8 | | -28.3 | | -6.5 | |
| ***Dried fruit*** [3] | | | | | | | | | | | | | | | | | | |
| Average dried fruit | 24.3 | 20.4 | 55.0 | 58.7 | 6.8 | 5.9 | 933 | 756 | | | 5 | 12 | 1.9 | 2.0 | 0.6 | 0.62 | 0.43 | 0.31 |
| Percentage difference from existing value, average | -16.0 | | 6.9 | | -12.8 | | -19.0 | | | | 166.7 | | 3.3 | | 10.0 | | -27.2 | |

| **Commodity** | **Moisture (g)** | | **Dietary fibre (g)** | | **B-carotene (μg)** | | **α-toco-pherol (mg)** | | **Ascorbic acid (mg)** | | **Folates, total (μg)** | | **Iron (mg)** | | **Zinc (mg)** | | **Copper (mg)** | |
|---|---|---|---|---|---|---|---|---|---|---|---|---|---|---|---|---|---|---|
| | **a** | **b** | **a** | **b** | **a** | **b** | **a** | **b** | **a** | **b** | **a** | **b** | **a** | **b** | **a** | **b** | **a** | **b** |
| ***Vegetables, raw*** [4] | | | | | | | | | | | | | | | | | | |
| Average | 88.7 | 89.1 | 3.2 | 3.0 | 1189 | 835 | 0.6 | 0.8 | 49 | 38 | 49 | 66 | 0.75 | 0.55 | 0.38 | 0.33 | 0.10 | 0.11 |
| Percent difference from existing value, average | 0.5 | | -7.3 | | -29.8 | | 27.8 | | -22.2 | | 35.2 | | -26.0 | | -13.6 | | 15.7 | |
| ***Nuts*** [5] | | | | | | | | | | | | | | | | | | |
| Average | 15.6 | 14.8 | 8.6 | 8.7 | 48 | 73 | 8.8 | 6.3 | 2 | 7 | 44 | 95 | 2.6 | 2.4 | 1.91 | 1.69 | 0.870 | 0.743 |
| Percent difference from existing value, average | -5.1 | | 0.9 | | 52.1 | | -28.4 | | 250 | | 116 | | -8.1 | | -11.5 | | -14.6 | |

[1] Values are only presented where both the old and new data included the analyte of interest

[2] Apple (red-skinned peeled, red-skinned unpeeled, Golden delicious unpeeled & Granny Smith unpeeled), apricot, banana (Cavendish & Lady Finger), blackberry, blueberry, raspberry, strawberry, cherry, custard apple, grapefruit (pink- & white-fleshed), grape (green & red), lemon, lime, lychee, mandarin, mango, melon (honeydew, rockmelon, watermelon), nectarine (white- & yellow-fleshed), orange, papaya (red- & yellow-fleshed), passionfruit, peach (white- & yellow-fleshed, pear (nashi, brown- & green-skinned), persimmon, pineapple, plum

[3] Apricot, currant, raisin, sultana and prune

[4] Avocado (Hass & Shepard), green bean, beetroot, bok choy, broccoli, Brussels sprout, cabbage (white & red/purple), capsicum (green & red), carrot, cauliflower, celery, chilli (red Thai style), cucumber (Lebanese), eggplant, kale, leek, lettuce (Cos), mushroom (common), onion (brown-, white- & red-skinned), parsnip, pea, potato, pumpkin, silverbeet, snow pea, spinach (baby), sweetcorn, sweet potato (orange)

[5] Almond, chestnut, macadamia, pistachio

**Table 2.** Variation in ascorbic acid levels within commodities, grouped by plant type. All values are per 100 g edible portion, fresh weight

| Commodity group | Includes | **Ascorbic acid content (mg)** | **Standard deviation (mg)** | **Relative standard deviation of mean (%)** |
|---|---|---|---|---|
| Apples | Green-, red- and yellow skinned apples, peeled and unpeeled | 4.7 | 3.1 | 66.0 |
| Citrus fruit | Pink- and red-grapefruit, lemon, lime, mandarin, orange | 37.6 | 9.0 | 23.7 |
| Melons | Honeydew melon, rockmelon, watermelon | 9.2 | 4.1 | 44.6 |
| Tropical fruit | Custard apple, mango, red- and yellow papaya, persimmon, pineapple, | 47.3 | 14.8 | 31.3 |
| Brassica vegetables | Bok choy, Broccoli, Brussels sprouts, red- and white-cabbage, cauliflower, kale | 70.2 | 17.4 | 24.8 |
| Roots and tubers | Potatoes, parsnip, sweet potatoes (orange, purple, white) | 11.9 | 4.0 | 33.6 |
| Leaf & stem green vegetables | Leek, silverbeet | 12.5 | 8.1 | 64.8 |
| Fruiting vegetables | Eggplant, pumpkin, zucchini | 9 | 3.2 | 35.6 |

**Supplementary Table 1.** Sampling details for analysis of the nutrient composition of Australian-grown horticultural commodities

| Sample name and description | Botanical name, alternate common names and varietal names | Primary samples, n | Purchase date |
|---|---|---|---|
| ***Fruits, fresh, raw*** | | | |
| Apple, red skinned, Pink Lady variety, unpeeled | *Malus pumila*. Cripps Pink variety. | 10 | April 2022 |
| Apple, red skinned, Pink Lady variety, peeled | *Malus pumila*. Cripps Pink variety. | 10 | April 2022 |
| Apple, green skinned, Granny Smith variety, unpeeled | *Malus pumila* | 10 | April 2022 |
| Apple, yellow skinned, Golden Delicious variety, unpeeled | *Malus pumila* | 9 | April 2022 |
| Apricot, flesh and skin | *Prunus armeniaca* | 10 | December 2021 |
| Banana, Cavendish variety, flesh | *Musa acuminata* | 10 | November 2021 |
| Banana, Lady Finger variety, flesh | *Musa acuminata.* Sugar banana. | 10 | November 2021 |
| Berry, blackberry | *Rubus* species | 10 | December 2021 |
| Berry, blueberry | Genus *Vaccinium* | 10 | December 2021 |
| Berry, raspberry | *Rubus* species | 10 | December 2021 |
| Berry, strawberry | Genus *Fragaria* | 9 | November 2021 |
| Cherry, flesh and skin | Genus *Prunus* | 10 | December 2021 |
| Custard Apple, flesh | *Annona squamosa x Annona cherimola.* African Pride and Pinks Mammoth are common varieties. | 9 | July 2021 |
| Grapefruit, pink fleshed, flesh | *Citrus x Paradisi.* Ruby (ruby blush) grapefruit. | 9 | July 2021 |
| Grapefruit, yellow fleshed, flesh | *Citrus x Paradisi.* Marsh is a common variety. | 7 | July 2021 |
| Grapes, green, flesh and skin | *Vitis vinifera* | 10 | January 2022 |
| Grapes, red. flesh and skin | *Vitis vinifera* | 10 | February 2022 |
| Lemon, flesh | *Citrus limon* | 8 | July 2021 |
| Lime, flesh | *Citrus aurantifolia*. Tahitian lime. | 8 | July 2021 |
| Lychee, flesh | *Litchi chinensis* | 10 | January 2022 |
| Mandarin, flesh | *Citrus reticulata*. Purchased varieties include Imperial, Murcott, Afourer | 9 (19 for folates) | July 2021 (all components) and May 2022 (folates only) |
| Mango, Kensington Pride variety, flesh | *Mangifera indica*. Bowen mango. | 9 | January 2022 |
| Melon, Honeydew variety, flesh | *Cucumis melo* | 10 | November 2021 |
| Melon, rockmelon variety, flesh | *Cucumis melo*. Canteloupe. | 10 | November 2021 |
| Melon, watermelon, flesh | *Citrulis lannatus* | 10 | November 2021 |
| Nectarine, white flesh, flesh and skin | *Prunus persica var. nucipersica* | 10 | December 2021 |
| Nectarine, yellow flesh, flesh and skin | *Prunus persica var. nucipersica* | 10 | December 2021 |
| Orange, flesh | *Citrus sinensis*. Mixture of Navel and Valencia types. Red-fleshed oranges not included. | 8 | July 2021 |
| Papaya, red fleshed, flesh | *Carica papaya* | 10 | June 2021 |
| Papaya, yellow fleshed, flesh | *Carica papaya.* Pawpaw. | 8 | June 2021 |
| Passionfruit, pulp | *Passiflora edulis* | 8 (except 18 for tocopherols) | October 2021 (and May 2022) |
| Peach, white fleshed, flesh and skin | *Prunus persica* | 10 | December 2021 |
| Peach, yellow fleshed, flesh and skin | *Prunus persica* | 10 | December 2021 |
| Pear, brown skinned, unpeeled | Genus *Pyrus* | 10 | March 2022 |
| Pear, green skinned, unpeeled | Genus *Pyrus*. Packham pear, Bartlett pear. | 10 | March 2022 |
| Pear, Nashi variety, unpeeled | *Pyrus pyrifolia.* Nashi, Asian pear. | 10 | March 2022 |
| Persimmon, flesh and skin | *Diospyros kaki*, may include sweet or astringent varieties | 6 | June 2021 |
| Pineapple, flesh, without core | *Ananus comosus* | 10 | October 2021 |
| Plum, pink, red or purple flesh and skin, flesh and skin | *Prunus domestica* | 10 | February 2022 |
| ***Fruits, dried*** | | | |
| Apricot, dried | *Prunus armeniaca* | 9 (3 and 6) | June 2021 and May 2022 |
| Grape, dried (currant) | *Vitis vinifera* | 7 | June 2021 |
| Grape, dried (raisin) | *Vitis vinifera* | 7 | June 2021 |
| Grape, dried (sun muscat) | *Vitis vinifera* | 10 | November 2021 |
| Grape, dried (sultana) | *Vitis vinifera* | 10 | June 2021 |
| Plum, dried (prune) | *Prunus domestica* | 9 | June 2021 |
| ***Vegetables, fresh*** | | | |
| Avocado, Hass variety, flesh | *Persea americana* | 8 | August 2021 |
| Avocado, Shepard variety, flesh | *Persea americana* | 10 | March 2022 |
| Beans, green | *Phaseolus vulgaris*. French beans, string beans. | 10 | December 2021 |
| Beetroot, root only, peeled | *Beta vulgaris*. Red-coloured root only. | 10 | September 2021 |
| Bok choy, leaves and core | *Brassica rapa subsp. Chinensis. Baby bok choy, Buk choy, Pak choy* | 8 | August 2021 |
| Broccoli, florets | *Brassica oleracea var italica.* Dark green in colour. | 8 | August 2021 |
| Brussels sprouts, whole | *Brassica oleracea var. gemmifera* | 12 | August 2021 (n=7) and May 2022 (n=5) |
| Cabbage, red or purple, leaves | *Brassica oleracea var capitata* | 8 | August 2021 |
| Cabbage, white, leaves | *Brassica oleracea var capitata*. Common cabbage. | 11 | August 2021 (n=6) and May 2022 (n=5) |
| Capsicum, green | *Capsicum annuum*. Bell pepper | 10 | January 2022 |
| Capsicum, red | *Capsicum annuum*. Bell pepper | 10 | January 2022 |
| Capsicum, yellow | *Capsicum annuum*. Bell pepper | 10 | January 2022 |
| Carrot, mature, unpeeled | *Daucus carota*. Orange coloured varieties. | 10 | September 2021 |
| Cauliflower, white, florets | *Brassica oleracea var. botrytis* | 8 | August 2021 |
| Celery, stalk | *Apium graveolens* | 10 | September 2021 |
| Chilli, red Thai style | *Capsicum annuum*. Long red chilli | 10 | January 2022 |
| Cucumber, Lebanese variety, unpeeled | *Cucumus sativus*. Dark green skin. | 10 | November 2021 |
| Cucumber, Lebanese variety, peeled | *Cucumus sativus*. White flesh. | 10 | November 2021 |
| Eggplant, cooked | *Solanum melangena*. Aubergine. | 10 | February 2022 |
| Eggplant, raw | *Solanum melangena*. Aubergine. | 10 | February 2022 |
| Kale, leaves | *Brassica oleracea var. sabellica.* Tuscan kale, curly kale, baby kale. | 8 | August 2021 |
| Leek, stem | *Allium porrum* | 10 | October 2021 |
| Lettuce, cos variety | *Lactuca sativa L. var. longifolia.* Romaine lettuce, green coloured. | 10 | February 2022 |
| Mushroom, common, with skin, cooked | *Agaricus bisporus* | 10 | September 2021 |
| Mushroom, common, with skin, raw | *Agaricus bisporus* | 10 | September 2021 |
| Onion, brown skin, peeled | *Allium cepa* | 10 | October 2021 |
| Onion, red skin, peeled | *Allium cepa* | 10 | October 2021 |
| Onion, white skin, peeled | *Allium cepa* | 12 | October 2021 (n=6) & May 2022 (n=6) |
| Parsnip, peeled | *Pastinaca sativa* | 10 | September 2021 |
| Peas, green | *Pisum sativum* | 8 | October 2021 |
| Potato, various types, unpeeled, raw | *Solanum tuberosum.* Includes potatoes sold as red-skinned, white-skinned or kipfler. | 10 | September 2021 |
| Pumpkin, flesh | *Cucurbita pepo.* Includes Kent, Butternut, Golden Nugget and Japanese style pumpkins. | 10 | June 2021 |
| Pumpkin, Butternut variety | *Cucurbita pepo* | 10 | January 2022 |
| Pumpkin, Kent variety | *Cucurbita pepo* | 10 | April 2022 |
| Silverbeet, leaves | *Beta vulgaris*. Chard. Dark green leaves. | 10 | October 2021 |
| Snow peas, pod and peas | *Pisum sativum* | 10 | October 2021 |
| Spinach, baby, leaves | *Spinacea oleracea* | 10 | October 2021 |
| Sweetcorn, kernels | *Zea mays* convar. *saccharata* var. *rugosa* | 10 | February 2022 |
| Sweet potato, orange fleshed, peeled | *Ipomoea batatas*. Kumara. | 8 (additional 10 for tocopherols) | August 2021 and April 2022 (tocopherols) |
| Sweet potato, orange fleshed, unpeeled | *Ipomoea batatas*. Kumara. | 10 | April 2022 |
| Sweet potato, purple fleshed | *Ipomoea batatas* | 7 | August 2021 |
| Sweet potato, white fleshed | *Ipomoea batatas*. Generally have brown skin | 9 | September 2021 |
| Zucchini, flesh and green skin | *Cucurbita pepo.* Courgette. | 10 | September 2021 |
| ***Nuts (unsalted)*** | | | |
| Almond, raw, with skin | *Prunus dulcis* | 10 | March 2022 |
| Chestnut, roasted, peeled | *Castanea sativa* | 10 | June 2021 |
| Macadamia, raw, without skin | *Macadamia integrifolia*. Queensland nut. | 10 | March 2022 |
| Pistachio, raw, with skin | *Pistacia vera* | 8 | March 2022 |

**Supplementary Table 2.** Analytical methods used for analysis of the nutrient composition of Australian-grown horticultural commodities, compared to methods used before 2000

| Nutrient | Units | Method 2021-22 | Method pre-2000 (where known)[1] |
|---|---|---|---|
| *Proximates* | | | |
| Moisture | g/100 g | Air or vacuum oven drying: AOAC 934.06, 964.22 (AOAC International, 1995), AS2300.1.1 (Standards Australia, 2008) | Same |
| Protein | g/100 g | Kjeldhal nitrogen based on AOAC 981.10, 920.152, 990.03, 920.87 (AOAC International, 2005) | Same |
| Fat | g/100 g | Mojonnier extraction AOAC 954.02,948.15,922.08 (AOAC International, 2005) | Same |
| Ash | g/100 g | Dry ashing. AOAC 923.03, 900.02 (AOAC International, 2005) | Same |
| Dietary fibre [a] | g/100 g | ANKOM dietary fibre analyser (ANKOM Technology), automating AOAC 2001.03 | Similar in principle, likely to be differences in steps such as digestion time, enzyme composition |
| Sugars | g/100 g | HPLC with refractive index detection, based on various methods including Sims (1995) | Similar, but likely to be different column packings and operating conditions |
| Starch [a] | g/100 g | HPLC as for sugars after enzymatic conversion of starch to glucose | Similar |
| Inulin | g/100 g | Enzyme gravimetric method AOAC 999.03 (AOAC International, 2000) | Not reported prior to around 2000 |
| Sorbitol | g/100 g | HPLC with refractive index detection, based on various methods including Sims (1995) | Not reported in Lim (1985) |
| *Organic acids* | | | |
| Malic, citric and lactic acids | g/100 g | In-house method based on Doyon, Gaudrea, St-Gelais, Beaulieu & Randall (1991) | Similar, but with different column packings and operating conditions |
| *Water-soluble vitamins* | | | |
| Ascorbic acid | mg/100 g | Total ascorbic acid, in-house HPLC method based on Brubacher, Müller-Mulot, & Southgate (1985) | AOAC microfluorometric method 43.061 - 43.067 (AOAC International, 1980) |
| Thiamin and Riboflavin | mg/100 g | European Standard EN 14122:2003/AC (European Committee for Standardization, 2003). Fluorescence detection | Similar method. AOAC, section 43.024 - 43.030 and section 43.039 - 43.042 with fluorometric detection (AOAC International, 1980) |
| Niacin | mg/100 g | HPLC with fluorescence detection, based on AOAC 43.045 (AOAC International, 2005) and Lahély, Bergaentzlé, & Hasselmann (1999) | AOAC section 43.044 - 43.046, using cyanogen bromide (AOAC International, 1980) |
| Pantothenic acid | mg/100 g | VitaFast® Pantothenic acid microtiter plate test AOAC-RI 100904 (r-biopharm, 2022a) | Not reported in Lim (1985). Analyses since around 1990 likely to have used a similar method to current method |
| Pyridoxine | mg/100 g | (Bergaentzlé, Arella, Bourguignon, & Hasselmann, 1995) | Not conducted in 1980s |
| Folates, total | µg/100 g | Microbiological assay using triple enzyme extraction, spectrophotometric measurement of turbidity | Single enzyme extraction, otherwise similar. Not conducted before 1990s |
| Biotin | µg/100 g | VitaFast® Biotin microtiter plate test AOAC-RI 101001 (r-biopharm, 2022b) or in-house LC/MS/MS method | Not reported in Lim (1985). Data generated in 1990s likely to have used a similar, but less developed, approach as the VitaFast test. |
| Tryptophan | mg/100 g | UPLC-PDA-MSMS following hydrolysis with NaOH | HPLC following hydrolysis with $Ba(OH)_2$ |
| *Fat-soluble vitamins* | | | |
| Alpha & beta carotene | µg/100 g | Reverse phase HPLC, UV detection, based on CRC Handbook of Chemistry and Physics (Weast, 1975) | Similar to current method but using Column chromatography, UV spectrophotometry |
| Cryptoxanthin, lutein and lycopene | µg/100 g | As for alpha and beta carotene and also Barba, Hurtado, Mata, Ruiz, & Tejada (2006) and Bushway (1985) | Column chromatography, UV spectrophotometry. Only cryptoxanthin reported prior to 2000 |
| Alpha, beta, gamma & delta tocopherols | mg/100 g | HPLC with fluorescence detection after saponification of the sample based on various publications including Shin & Godber (1994) | Not reported prior to about 2000. Likely to be similar method since that time |
| *Minerals* | | | |
| All minerals except iodine | mg or µg /100 g | Acid digest, ICP-MS/ICP-OES: NMI in-house method based on: (US Environmental Protection Agency, 1994, 1996) | Atomic absorption spectroscopy to mid-1980s, then ICP-MS or ICP-OAS.<br>Not all minerals reported before 2000 (e.g. Al, Mn) |
| Iodine | µg/100 g | Alkaline extraction, ICP MS. NMI in-house method based on: (Fecher, Goldmann, & Nagengast, 1998) | Not reported prior to 2000 |
| Fatty acids | % of total fatty acids | Gas-liquid chromatography of methyl esters using flame ionisation detection | Similar, but with different column packings and operating conditions, more limited standards of identification |

All analyses conducted in Australia by NMI (Port Melbourne), except [a] AEGIC (North Ryde), [b] BVAQ (Melbourne) and [c] PathWest (Perth)
AEGIC, Australian Export Grains Innovation Centre; AOAC-RI, Association of Official Analytical Chemists Research Institute; BVAC, Bureau Veritas AsureQuality; ICP MS, inductively coupled plasma mass spectrometry; LOR, limit of reporting; NMI, National Measurement Institute of Australia
[1]Much of data from pre-2000 was privately reported to the now FSANZ and not published and as a result, some details on methods of analysis have been lost

**Supplementary Table 2 References**